\documentclass[nofootinbib,preprint,superscriptaddress,showpacs,amsmath,amssymb,aps,pra,showkeys]{revtex4-2}

\usepackage{amsmath,amsfonts,amssymb}
\usepackage{graphicx}
\usepackage{xcolor}
\usepackage{float}

\usepackage{calc}
\usepackage[normalem]{ulem}

\usepackage{accents}

\begin{document}

\title{Transfer learning, alternative approaches, and visualization of a convolutional neural network for retrieval of the internuclear distance in a molecule from photoelectron momentum distributions}

\author{N. I. Shvetsov-Shilovski}
\email{n79@narod.ru}
\affiliation{Institut f\"{u}r Theoretische Physik, Leibniz Universit\"{a}t Hannover, 30167 Hannover, Germany}

\author{M. Lein}
\affiliation{Institut f\"{u}r Theoretische Physik, Leibniz Universit\"{a}t Hannover, 30167 Hannover, Germany}

\date{\today}

\begin{abstract}
We investigate the application of deep learning to the retrieval of the internuclear distance in the two-dimensional H$_2^{+}$ molecule from the momentum distribution of photoelectrons produced by strong-field ionization. We study the effect of the carrier-envelope phase on the prediction of the internuclear distance with a convolutional neural network. We apply the transfer learning technique to make our convolutional neural network applicable to distributions obtained for parameters outside the ranges of the training data. The convolutional neural network is compared with alternative approaches to this problem, including the direct comparison of momentum distributions, support-vector machines, and decision trees. These alternative methods are found to possess very limited transferability. Finally, we use the occlusion-sensitivity technique to extract the features that allow a neural network to take its decisions.\\
\end{abstract}



\maketitle

\newpage

\section{Introduction} 

Machine learning focuses on the development of algorithms and methods that are able to learn, i.e., to use data in order to improve automatically through experience \cite{Mitchell1997}. Methods of machine learning are presently widely used in almost all branches of modern science. Strong-field, ultrafast, and attosecond physics that studies the nonlinear processes originating from interaction of strong laser pulses with atoms and molecules \cite{BeckerRev2002,MilosevicRev2003,FaisalRev2005,FariaRev2011,Graefe2016,Lin2018} is no exception. The prediction of the flux of high-order harmonics \cite{Gherman2018}, the prediction of the ground-state energy of an electron confined by a two-dimensional (2D) potential \cite{Mills2017}, the retrieval of the intensity and the carrier-envelope phase (CEP) of ultrashort laser pulses from frequency-resolved optical gating traces \cite{Zahavy2018} and dispersion scan traces \cite{Kleinert2019}, the efficient implementation of the trajectory-based Coulomb-corrected strong-field approximation \cite{Yan2010,Yan2012} by using of a deep neural network \cite{Yang2020} are examples of machine-learning applications in strong-field physics and related areas. More recently, convolutional neural networks have been used to predict high-order harmonic spectra for model di- and triatomic molecules when the laser intensity, internuclear distance, and orientation of the molecule were given and vice versa, to retrieve molecular parameters from a given high-order harmonic spectrum, see Ref.~\cite{Lytova2022}. Recently a convolutional neural network was used to retrieve the geometric structure of gas-phase molecules from experimentally measured laser-induced electron diffraction images \cite{Biegert2021}. Furthermore, machine learning was applied to retrieve the internuclear distance in a molecule from a given photoelectron momentum distribution produced by a strong laser pulse \cite{Shvetsov2022}. The problems considered in Refs.~\cite{Biegert2021,Shvetsov2022} are important for the development of tools aimed at time-resolved molecular imaging, i.e., visualization of molecular dynamics in real time. 

Time-resolved molecular imaging requires high resolution in both space and time. Indeed, molecular transformations occur due to motion of atoms on the angstrom scale. Simultaneously, a chemical reaction has a typical duration of less than a picosecond. It would be desirable to apply time-resolved molecular imaging to large molecules, e.g., biomolecules. Eventually, these techniques should allow to study not only the dynamics of the atomic nuclei, but also the electronic dynamics. Important methods for time-resolved molecular imaging include, for example, optical pump-probe spectroscopy, time-resolved electron and X-ray diffraction, and ultrafast X-ray spectroscopy ~\cite{Agostini2016}. 

The advances in laser technologies over the last decades, especially the emergence of table-top intense optical laser systems and progress in the development of free-electron lasers, have offered possibilities of new methods for real-time molecular imaging. Among these new methods based on strong-field phenomena are laser-induced Coulomb explosion imaging \cite{Frasinski1987,Cornaggia1991,Posthumus1995,Cornaggia1995}, laser-assisted electron diffraction \cite{Kanya2010,Morimoto2014}, high-order harmonic orbital tomography \cite{Itatani2004,Haessler2010}, laser-induced electron diffraction \cite{Lein2002,Meckel2008,Blaga2012,Pullen2015}, and strong-field photoelectron holography \cite{Huismans2011}. Laser-induced electron diffraction and strong-field photoelectron holography rely on the analysis of momentum distributions of electrons removed by a strong laser pulse. At present, these methods are beginning to be successfully applied to dynamical systems, see, e.g., Refs.~\cite{Haertelt2016,Walt2017}. Therefore, in the near future we can expect experiments aimed at obtaining information about the nuclear motion in a molecule interacting with a strong laser pulse from photoelectron momentum distributions (PMDs). The interpretation of such an experiment requires insight into the relation between momentum distribution and internuclear distance at fixed nuclei.  

In Ref.~\cite{Shvetsov2022} a convolutional neural network (CNN) was used to retrieve the internuclear distance of a 2D model H$_2^{+}$ molecule from photoelectron momentum distributions generated by a strong few-cycle laser pulse. The momentum distributions were calculated from the direct numerical solution of the time-dependent Schr\"{o}dinger equation (TDSE). It was shown that a CNN trained on a relatively small number of electron momentum distributions predicts the internuclear distance with a mean absolute error (MAE) below $0.1$~a.u. Furthermore, a neural network was able to retrieve both the internuclear distance and the laser intensity from a given photoelectron momentum distribution. Finally, the effect of focal averaging was studied in Ref.~\cite{Shvetsov2022}. It was shown that a CNN trained on focal averaged momentum distributions also performs well in the retrieval of the internuclear distance. 

A number of important questions concerning deep learning for the molecular-structure retrieval remain to be studied. First, it is important to study the effect of the CEP, since the variation of the CEP of a few-cycle pulse affects the PMDs significantly. Second, the CNN presented in Ref.~\cite{Shvetsov2022} shows limited transferability, i.e., it may fail for PMDs corresponding to parameters beyond the ranges of the training data. The transferability problem can be tackled with the transfer learning technique \cite{Goodfellow}. 

Furthermore, it is interesting to compare the CNN with alternative methods. The simplest possible method is the direct comparison of a given PMD with a precalculated set of PMDs obtained for various internuclear distances \cite{Shvetsov2022}. Support vector machines (see, e.g., Refs.~\cite{Vapnik1995,VapnikBook,ScholkopfBook,SteinwartBook}) and decision trees \cite{BreimanBook,Quinlan1986,Breiman1996} are further alternatives. Both methods, combined with the histogram of oriented gradients  \cite{Dalal2005}, were extensively used for object detection and image comparison before CNNs became widespread. Finally, since a CNN is a ``blackbox", it is interesting to understand how the CNN of Ref.~\cite{Shvetsov2022} takes its decisions, i.e., to have ``a look under the hood" of the neural network. Although this is a difficult task, a number of so-called visualization and explanation methods for deep neural networks have been developed Refs.~\cite{Atefeh2021,Linardatos2021,Ayyar2021}. These methods can potentially uncover the features of the PMDs that are used by the neural network to recognize the internuclear distance.

In this paper we address the above-mentioned questions. We apply the CNN to PMDs obtained for three variable parameters: internuclear distance, laser intensity, and CEP. Thus, we study the effect of the CEP on the reconstruction of the internuclear distance. Using the transfer learning technique, we achieve transferability of the CNN and we perform a number of transferability tests. We then compare the machine-learning approach with alternative methods. For the purpose of direct comparison we apply not only the mean squared pixel-wise error, but also the histogram of oriented gradients (HOG) and the scale-invariant feature transform (SIFT) algorithm proposed by D.~Lowe \cite{Lowe1999}. The image descriptors obtained from the SIFT algorithm are invariant with respect to uniform scaling and orientation changes. They are also partially invariant with respect to affine distortions \cite{Lowe1999}. Finally, by applying visualization methods, we extract the features that allow the CNN to classify PMDs by internuclear distance.

The paper is organized as follows. In Sec.~II we briefly review the architecture of the CNN and the method used for the solution of the TDSE. 
In Sec.~III we study the effect of the CEP on the retrieval of the internuclear distance. In Sec.~IV we use the transfer learning to make the CNN applicable beyond the parameter ranges of the training data set. We compare the application of the CNN to alternative methods in Sec.~V, and in Sec.~VI we apply the visualization methods. The conclusions are given in Sec.~VII. Atomic units are used throughout the paper unless indicated otherwise. 

\section{Model}

The architecture of the neural network that we use for prediction of the internuclear distance was discussed in Ref.~\cite{Shvetsov2022}. We repeat here the most important points to make the presentation self-contained. The same applies to our method for the solution of the 2D TDSE. 

\subsection{Architecture and application of convolutional neural network}

The architecture of a neural network should be consistent with the format of the data used for training. For the problem at hand the training data are pairs of electron momentum distributions and the corresponding internuclear distances. In terms of machine learning, the distributions and the internuclear distances are the images and labels, respectively. Since we consider ionization of the molecule by a linearly polarized laser pulse, it is natural to assume the aspect ratio of every image as $2:1$ \cite{Shvetsov2022}.   

The images used by the CNN are preprocessed as follows. We calculate the decimal logarithm of the normalized PMD, i.e., $W=\text{log}_{10}\left(\text{PMD}/\text{PMD}_{\text{max}}\right)$, where $\text{PMD}_{\text{max}}$ is the absolute maximum of the distribution, and we set $W=-5$ instead of all values smaller than $-5$. Therefore, we account not only for the low-energy part of the distribution, but also the beginning of its high-energy part. The low-energy part is formed by the electrons that arrive at the detector without recolliding with their parent ions. These electrons are referred to as direct electrons. The direct electrons have energies below $2U_{p}$ (i.e., momenta below $2\sqrt{U_{p}}$), where $U_p=F^2/4\omega^2$ is the ponderomotive energy. In contrast to this, the high-energy part of the PMD arises due the rescattered electrons that are driven back to their parent ions by the laser field and rescatter off them by large angles. We select a rectangular part of the image that contains all values of W exceeding $-5$. By using bicubic interpolation, we downsize this rectangular part of the PMD to the size of $256\times128$ pixels, in order to avoid large matrices, which cause heavy computational costs and slow down the training process. We then rescale all the elements of the resulting matrix to map the maximum value to 255 and the minimum value to zero. The resulting images are given to the neural network.

\begin{figure}[h!]
\begin{center}
\includegraphics[width=.75\textwidth, trim=20 20 0 0]{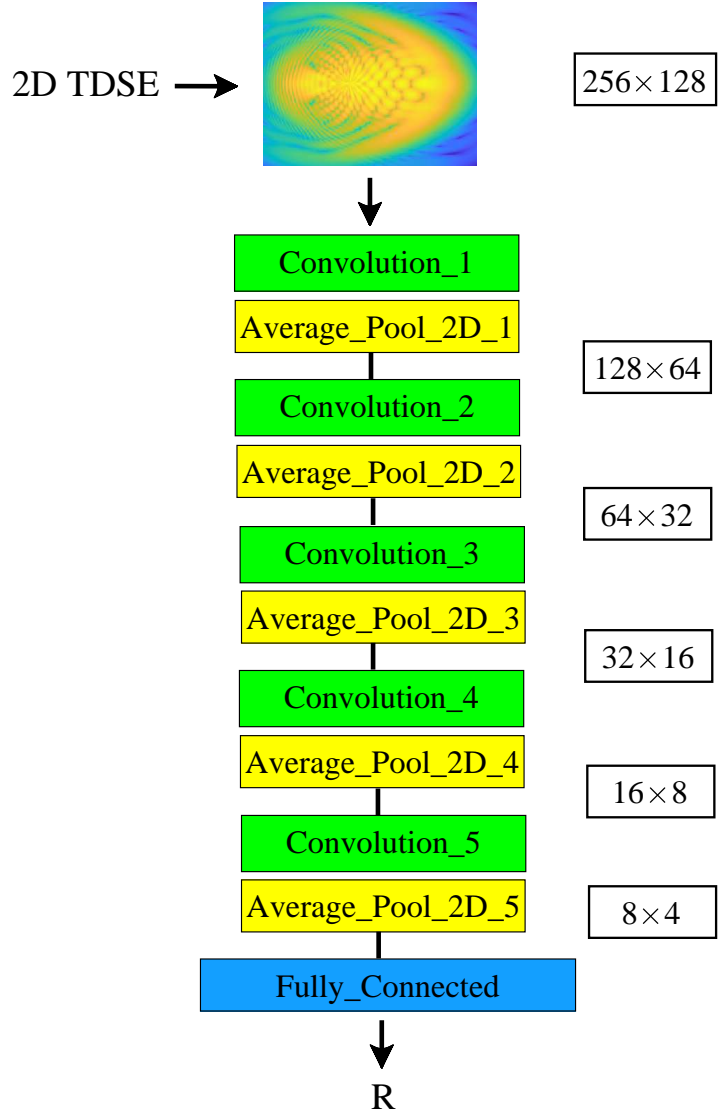} 
\end{center}
\caption{The architecture of the neural network used for retrieval of the internuclear distance $R$. The sizes of the image after each average pooling layer are indicated on the right side.}
\label{fig1}
\end{figure}

The architecture of the deep neural network is shown in Fig.~\ref{fig1}. The neural network consists of 5 pairs of nonreducing convolutional layers and reducing average pooling layers. We have found that using a smaller number of these pairs worsens the performance of the CNN. Simultaneously, an increased number does not lead to a significant improvement of the results. Each of the convolutional layers consists of 32 filters, and the size of each filter is $3\times3$ pixels. The padding needed to make the size of the output of each convolutional layer equal to the size of its input is to be calculated and used by the software in the training process. A convolutional layer convolves its input and produces new images (so-called feature maps ~\cite{KimBook}). The number of the feature maps is equal to the number of filters. The elements of the filter matrices are the main trainable parameters of the CNN. Following the convolution of their input images, the convolutional layers apply the rectified linear unit (ReLU) activation function, i.e., the piecewise linear function defined as $\text{ReLU}\left(x\right)=\max\left(0,x\right)$. 

The average pooling layers reduce the image size. These layers divide their input images into pooling regions and average the images over each pooling region. The size of the pooling regions is chosen to be $2\times2$ pixels, and therefore each of the average pooling layer reduces the image length and height by a factor of two. The output of the last average pooling layer is given to the dropout layer (not shown in Fig.~\ref{fig1}). The dropout layer randomly substitutes some fraction of its input values by zeros. In our case this fraction is chosen to be equal to $20$~\%. The presence of the dropout layer helps to prevent overfitting, i.e., the situation in which the CNN learns too many details of the training data including the noise and, as a result, performs poorly on data it has not seen before~\cite{TraskBook}. The dropout layer is connected to the last layer of our CNN, i.e., a fully connected layer. We use the CNN to solve the regression problem, not a classification problem. Indeed, we train a neural network to retrieve the internuclear distance $R$ from a given electron momentum distribution. Therefore, the fully connected layer calculates the output of the neural network – the internuclear distance $R$ (possibly along with any other value of interest). If only $R$ is to be calculated, the fully connected layer multiplies its input (column matrix) by a weight (row matrix) and adds a bias value. The neural network is implemented using the MATLAB package \cite{Matlab}.  

In the process of training, we minimize the loss function – the measure of deviation between predictions of the CNN and the known labels (internuclear distances) of the training set. The mean squared error (MSE) is used as the measure of the deviation. For minimization we apply the stochastic gradient descent optimizer. We split the training data into mini-batches, each consisting of $30$ images, and use one mini-batch for each training iteration. We start the training process with the learning rate $l_r=10^{-3}$ and decrease the rate by a factor of $10$ after $20$ training epochs. We find that about $30$ epochs are enough for convergence of the loss function. In order to ensure that each PMD creates an unbiased change in the CNN, we shuffle the training data before each training epoch. 

\subsection{Numerical solution of time-dependent Schr\"odinger equation}

We perform our TDSE simulations for a few-cycle linearly polarized laser pulse that is defined through the vector potential:
\begin{equation}
\label{vecpot}
\vec{A}\left(t\right)=\left(-1\right)^{n_p}\frac{F_0}{\omega}\sin^2\left(\frac{\omega t}{2n_p}\right)\sin\left(\omega t+\varphi\right)\vec{e}_{x}.
\end{equation}
Here $F_0$ is the field amplitude, $\omega$ is the laser frequency, $n_p$ is the number of optical cycles within the pulse, $\varphi$ is the CEP, and $\vec{e}_{x}$ is the unit vector in the direction of the $x$ axis (polarization direction). The laser pulse defined by Eq.~(\ref{vecpot}) is present between $t=0$ and $t=\left(2\pi/\omega\right)\cdot n_{p}$. The electric field is to be calculated from the vector potential (\ref{vecpot}) as $\vec{F}\left(t\right)=-d\vec{A}/dt$.  

The velocity gauge TDSE for the 2D H$_{2}^{+}$ molecular ion reads as
\begin{equation}
\label{2d_tdse}
i\frac{\partial}{\partial t}\Psi\left(x,y,t\right)=\left\{-\frac{1}{2}\left(\frac{\partial}{\partial x^2}+\frac{\partial}{\partial y^2}\right)-iA_{x}\left(t\right)\frac{\partial}{\partial x}+V\left(x,y\right)\right\}\Psi\left(x,y,t\right).
\end{equation}
Here $\Psi\left(x,y,t\right)$ is the wave function and 
\begin{eqnarray}
\label{potential}
V\left(x,y\right)=&-&\frac{1}{\sqrt{\left(x-\frac{1}{2}R\cos\alpha\right)^2+\left(y-\frac{1}{2}R\sin\alpha\right)^2+a}}\nonumber \\
                  &-&\frac{1}{\sqrt{\left(x+\frac{1}{2}R\cos\alpha\right)^2+\left(y+\frac{1}{2}R\sin\alpha\right)^2+a}}.
\end{eqnarray}
is the soft-core binding potential of the model H$_{2}^{+}$ molecule in the approximation of frozen nuclei. In Eq.~(\ref{potential}), $a$ is the soft-core parameter, $R$ is the internuclear distance, and $\alpha$ is the angle between the molecular axis and the polarization direction (orientation angle). We solve the TDSE (\ref{2d_tdse}) by applying the Feit-Fleck-Steiger split-operator method, see Ref.~\cite{Feit1982}. We use imaginary-time propagation to obtain the wave function of the ground state. 

The computational grid extending over $x\in\left[-400,400\right]$~a.u. and $y\in\left[-200,200\right]$~a.u. is centered at the origin $\left(x=0,y=0\right)$. Our grids in $x$ and $y$ directions consist of $4096$ and $2048$ points, respectively. Therefore, we use equal grid spacings for both directions: $\Delta x=\Delta y\approx0.1953$~a.u. The TDSE is propagated from $t=0$ (beginning of the laser pulse) to $t=4t_{f}$ with the time step $\Delta t=0.0184$~a.u. Absorbing boundaries are used to prevent unphysical reflections from the boundary of the computational box. More specifically, at every step of the time propagation the wave function is multiplied by the mask: 
\begin{equation}
M(x,y) = 
 \begin{cases}
   1 &\text{for $r\leq r_b$}\\
   \exp\left[-\beta\left(r-r_b\right)^2\right] & \text{for $r > r_b$}
 \end{cases},
\end{equation}
where $\beta=10^{-4}$, $r=\sqrt{x^2+y^2}$, and $r_b=150$~a.u. \cite{Shvetsov2022}. We obtain the electron momentum distributions by using the mask method, see Refs.~\cite{Lein2002,Tong2006}. 

In order to obtain the training and validation sets of PMDs we solve the TDSE (\ref{2d_tdse}) for $N$ randomly chosen internuclear distances and peak laser intensities: $R_{k}\in\left[1.0,8.0\right]$~a.u. and $I_{k}\in\left[1.0,4.0\right]\times10^{14}$ W/cm$^2$, where $k=1,...,N$ \cite{Shvetsov2022}. We also calculate focal volume averaged momentum distributions:
\begin{equation}
\label{aver}
\frac{dP}{d^{3}\vec{k}}=\int_{0}^{I_0}\frac{dP\left(I\right)}{d^{3}\vec{k}}\left(-\frac{\partial V}{\partial I}\right)dI,
\end{equation}
where $I_0$ is the peak laser intensity, $dP\left(I\right)/d^{3}\vec{k}$ is the PMD for a fixed intensity $I$, and $-\left(\partial V/ \partial I\right)dI$ is the focal volume element corresponding to the intensity range between $I$ and $I+dI$, see Ref.~\cite{Morishita2007}. Here we assume a Lorentzian distribution of the laser intensity along the polarization direction, and a Gaussian intensity distribution in the orthogonal direction, see, e.g., Refs.~\cite{SiegmanBook,Augst1991,Lin2018}. For such a beam the focal volume element reads as
\begin{equation}
\label{focvol}
\left(-\frac{\partial V}{\partial I}\right)dI\sim\frac{I_0}{I}\left(\frac{I_0}{I}+2\right)\sqrt{\frac{I_0}{I}-1}~dI.
\end{equation}
Obviously, the calculation of focal-volume averaged momentum distributions is a computationally demanding task. Indeed, a number of the solutions of the TDSE~(\ref{2d_tdse}) for different intensities $0<I<I_0$ is required to calculate the integral (\ref{aver}). We note that only about $100$ focal-volume averaged PMDs corresponding to different internuclear distances and laser intensities are needed for successful application of 2D interpolation on an irregular grid (see, e.g., Ref.~\cite{Press2007}) in the $\left(R,I_{0}\right)$ plane, see Ref.~\cite{Shvetsov2022}. This facilitates the calculation of large sets of focal-volume averaged momentum distributions. The interpolation is justified by the fact that focal averaged PMDs are rather smooth functions of both momentum components. 

\section{Effect of the carrier-envelope phase of the laser pulse}

\begin{figure}
\begin{center}
\includegraphics[width=.80\textwidth, trim=0 0 20 0]{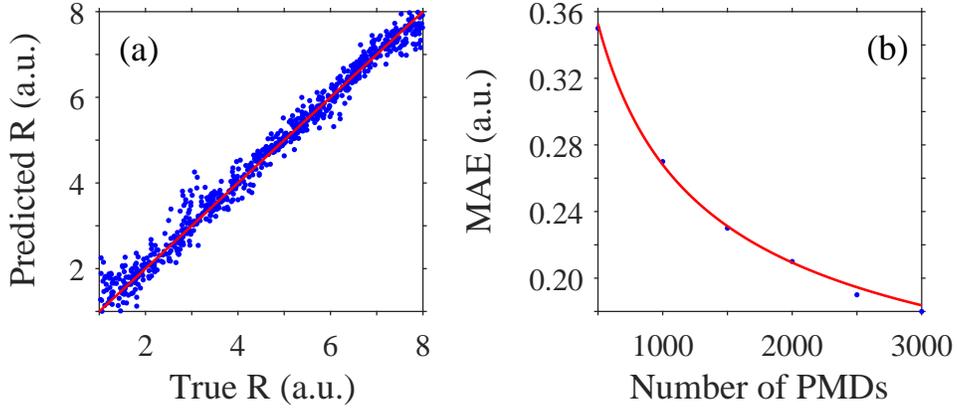} 
\end{center}
\caption{(a) Plot of predicted vs true internuclear distances illustrating the performance of the CNN for the distributions that depend on the three parameters:  laser intensity, internuclear distance, and CEP. (b) The MAE for the internuclear distance retrieved with the neural network as a function of number of images used for training. The blue dots and the red curve correspond to the points obtained in numerical experiments and the fit with a rational function (see text), respectively.}
\label{fig2}
\end{figure}

We first study the effect of the CEP on the retrieval of the internuclear distance using the neural network. To this end, we calculate $N=3000$ electron momentum distributions for $N$ random CEPs $\varphi_{k} \left(k=1,...,N\right)$. The corresponding internuclear distances $R_{k}$ and peak laser intensities $I_{k}$ are also chosen randomly in the ranges $\left[1.0,8.0\right]$~a.u. and $\left[1.0,4.0\right]\times10^{14}$~W/cm$^2$, respectively. Thus, the distributions of the new data set depend on the three random parameters: $R$, $I$, and $\varphi$. We then split this set of the PMDs into the training and test sets in the ratio $0.75:0.25$. The $2250$ distributions of the training set are used to train a neural network aimed at retrieval of the internuclear distance. This neural network is tested on the $750$ PMDs of the test set. We find that the internuclear distance is predicted with an MAE of $0.18$~a.u, see Fig.~\ref{fig2}(a). This should be compared with the MAE of $0.07$~a.u. obtained for the CNN trained on the distributions depending on only two parameters: internuclear distance and peak laser intensity. 

Then the following question arises: how many distributions depending on all the three parameters $\left(R,I,\varphi\right)$ are needed so that the network is able to predict the distance $R$ with a MAE less than a certain value., e.g., $0.1$~a.u.? In order to answer this question, we train the CNN using different numbers of electron momentum distributions ($N=500$, $1000$, $1500$, $2000$, and $2500$) and calculate the corresponding MAEs. Using this data, we fit the MAE as a function of $N$: $\text{MAE}=aN^{b}+c$, where $a=7.241$, $b=-0.526$, and $c=0.075$, see Fig.~\ref{fig2}(b). By extrapolating this fit, we find that about $50000$ images are needed to achieve an MAE less than $0.1$~a.u. We note that this result is similar to the outcomes of the study of Ref.~\cite{Lytova2022}: about $30000$ training samples that depend on three different parameters (internuclear distance, peak laser intensity, and the angle between the molecular axis and the polarization direction) were needed to predict the dipole acceleration with high accuracy \cite{Lytova2022}.  

\section{Transferability of the convolutional neural network}

We recall that the first neural network of Ref.~\cite{Shvetsov2022} is trained on a set of distributions calculated for random internuclear distances and fixed (not focal averaged) laser intensities. This CNN shows only limited transferability. We aim to develop new neural networks capable of predicting internuclear distances correctly even for PMDs calculated at parameters that are beyond the corresponding parameter ranges of the training data. To this end, we apply the transfer learning technique to the CNN of Ref.~\cite{Shvetsov2022}. We freeze all the weights of the original CNN except for those of the layers close to the output layer. Then this pre-trained CNN is further trained using a small data set that is outside of the initial training data space. In this way, we modify the CNN to make it applicable to (i) focal volume averaged PMDs, (ii) larger internuclear distances $8.0<R<12.0$~a.u., (iii) nonzero angles between the molecular axis and polarization direction, and (iv) nonzero CEPs.

\begin{figure}
\begin{center}
\includegraphics[width=.80\textwidth, trim=0 0 10 5]{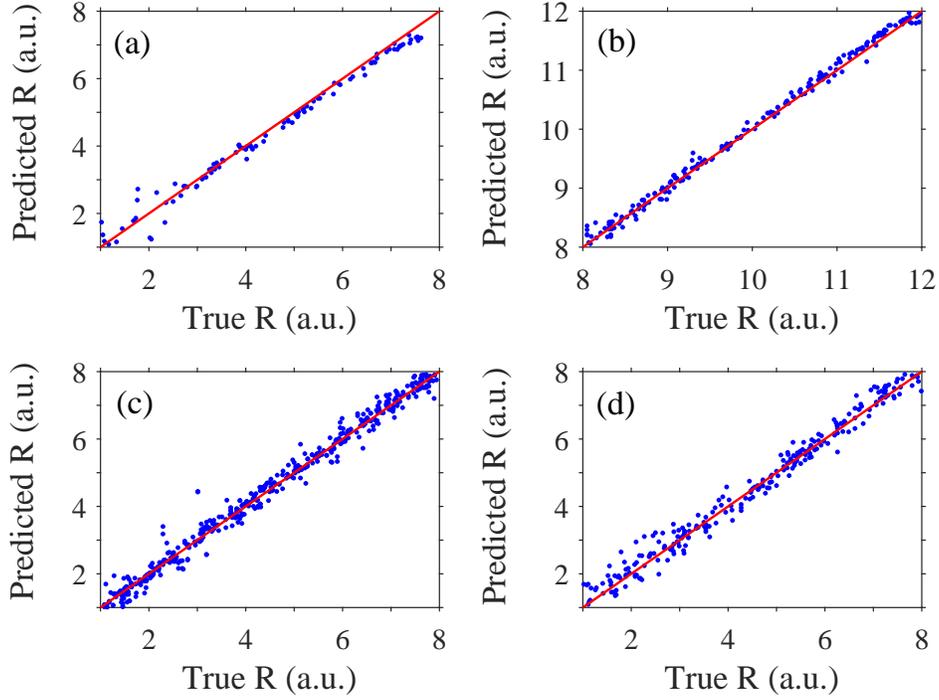} 
\end{center}
\caption{Plots of predicted vs. true internuclear distances illustrating the performance of the initial CNN [see Sec. II~(a)] after application of the transfer learning technique. Panels (a), (b), (c), and (d) correspond to the neural network receiving focal-volume averaged momentum distributions, distributions obtained for $8.0<R<12.0$~a.u., distributions calculated for different orientation angles $0^{\circ}<\alpha<180^{\circ}$, and distributions obtained for fluctuating CEP, respectively.}
\label{fig3}
\end{figure}

We begin with the momentum distributions averaged over the focal volume. As in Ref.~\cite{Shvetsov2022}, we use 2D interpolation on an irregular grid to obtain a set of $N=1000$ averaged momentum distributions from a smaller set consisting of only $N_a=100$ focal-volume averaged PMDs. We freeze the weights of the first four convolutional layers and retrain the fifth convolutional layer and the fully connected layer. The training is performed for mini-batches consisting of $30$ images with the learning rate $10^{-2}$. We find that the set of $N=750$ focal averaged PMDs is enough to achieve the MAE $0.15$ a.u. for the internuclear distance $R$, see Fig.~\ref{fig3}(a). It should be stressed that this MAE was obtained on the independent test set of calculated (not interpolated) focal volume averaged PMDs. A CNN trained on $6000$ focal volume averaged PMDs provides the MAE $0.14$ a.u. \cite{Shvetsov2022}. About $5$ minutes are needed to train the new CNN using $6000$ focal averaged PMDs, whereas the application of the transfer learning technique with $750$ averaged momentum distributions requires $35$-$40$ seconds. This applies to a modern PC using a graphic processing unit. Thus, we conclude that for focal averaged PMDs the transfer learning technique has one, but very important advantage compared to training of a new CNN from the ground up: a smaller training set is needed to achieve the same absolute error. Since the calculation of the training and validation data sets is usually computationally demanding, the application of transfer learning can reduce the computational costs substantially.

We then apply transfer learning to make our CNN applicable to distributions obtained for larger internuclear distances. To this end, we use a set of PMDs obtained for $8.0<R<12.0$~a.u. Here we fix the first three convolutional layers and retrain the neural network with the learning rate $10^{-3}$, since in the case of larger internuclear distances this provides better result as compared to the freezing of the first four convolutional layers and the use of the learning rate $0.01$. Only about $600$ images are needed to obtain the MAE $0.06$ a.u. for $R$, see Fig.~\ref{fig3}(b). We note that a new CNN trained on the same set of $600$ distributions with large internuclear distances provides the MAE of $0.11$~a.u. Thus, a set of $600$ distributions is not sufficient to achieve the MAE less than $0.1$ by training a new CNN. The transfer learning technique provides better results for the same number of distributions. This could be expected, since more trainable parameters are to be optimized in the training of a new neural network compared to the retraining of only $2$-$3$ layers of the CNN as required by the transfer learning technique.

\begin{figure}
\begin{center}
\includegraphics[width=.45\textwidth, trim=10 10 0 0]{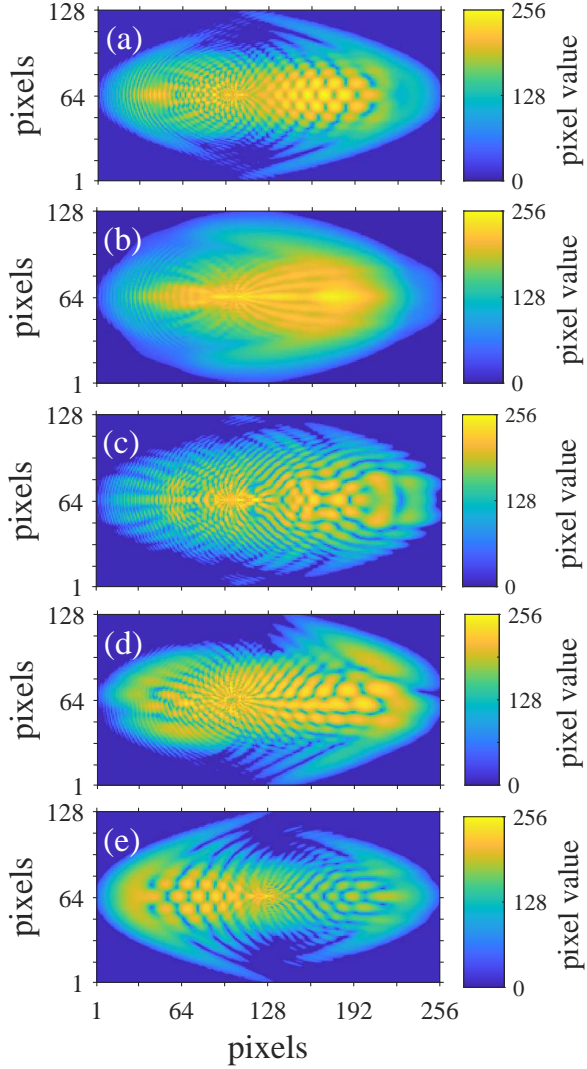} 
\end{center}
\caption{Electron momentum distributions for ionization of the 2D H$_2^{+}$ molecule as they are seen by the neural network, i.e., as $256\times 128$ matrices with element values between $0$ and $255$. The distributions are obtained from the solution of the TDSE. The laser pulse duration is $n_p=2$ cycles, the wavelength is $800$~nm, and the peak laser intensity is $3.9\times10^{14}$ W/cm$^2$. Panel (a) corresponds to the internuclear distance $R=2.0$~a.u, the orientation angle $\alpha=0^{\circ}$, and the CEP $\varphi=0$. Panel (b) shows the focal-volume averaged momentum distribution calculated for the parameters of panel (a). Panel (c) corresponds to the internuclear distance $R=10.9$~a.u., with the other parameters as in panel (a). Panel (d) displays the distribution calculated for $\alpha=158^{\circ}$ with the rest of the parameters as in panel (a). Panel (e) corresponds to the CEP $\varphi=4.50$~rad with the rest of the parameters as in panel (a).}
\label{fig4}
\end{figure}

It should be noted that focal averaging and larger internuclear distances are relatively easy cases for transfer learning. Indeed, the focal volume averaged PMDs and those corresponding to larger internuclear distances resemble (in terms of the symmetry and the positions of their maxima) the distributions calculated for fixed laser intensities and smaller values of $R$ [compare Fig.~\ref{fig4}(a) with Figs.~\ref{fig4}(b) and \ref{fig4}(c)]. The situation changes for the PMDs corresponding to nonzero angles between the molecular axis and the laser polarization direction, see Fig.~\ref{fig4}(a) and Fig.~\ref{fig4}(d). It is seen that the distribution of Fig.~\ref{fig4}(d) is substantially deformed as to Fig.~\ref{fig4}.(a), and this deformation is rather complex. As a result, the transfer learning technique performs worse than in the two previous cases. Indeed, the application of transfer learning using $750$ PMDs calculated for random angles $0^{\circ}\leq\alpha\leq 180^{\circ}$ leads to the MAE $0.17$~a.u., see Fig.~\ref{fig3}(c). Here we again fix the first three convolutional layers and choose the learning rate $l_{r}=10^{-3}$. This result should be compared with the MAE of 0.25 provided by the a CNN trained on the same set of $750$ distributions with nonzero orientation angles. Therefore, transfer learning provides a significant advantage. On the other hand, when $2250$ PMDs obtained for nonzero angles are used for training of the new neural network, the corresponding MAE is equal to $0.18$~a.u., which is close to the transfer learning result achieved with $750$ distributions only. Nevertheless, we made an attempt to improve the outcomes of transfer learning without calculating new momentum distributions. To this end, we have augmented the data set used for transfer learning by rotating the available PMDs by small angles and reflecting them with respect to the polarization direction. However, this does not reduce the corresponding MAE substantially. The same applies to a modified approach replacing the last 2-3 layers by new untrained ones and subsequent training of them.

Finally, we apply transfer learning to the case of nonzero CEP. Freezing the first three layers of the neural network, using a set of $750$ distributions calculated for random CEPs, and choosing the learning rate $10^{-2}$, we obtain the MAE $0.23$~a.u., see Fig.~\ref{fig3}(d). This result should be compared with the MAE of $0.31$~a.u. obtained for a new CNN trained on the same set of $750$ momentum distributions, as well as with the MAE $0.18$~a.u., corresponding to the training of a new CNN using $2250$ PMDs with random CEPs, see Sec.~III. We conclude that transfer learning provides some gain even in the case of varying CEP, which judging from the strong CEP dependence of the PMDs, would seem to be a difficult case [cf. Figs.~\ref{fig4}(a) and \ref{fig4}(e)].

\section{Alternative approaches for the retrieval of the internuclear distance}

\begin{figure}
\begin{center}
\includegraphics[width=.95\textwidth, trim=0 0 10 10]{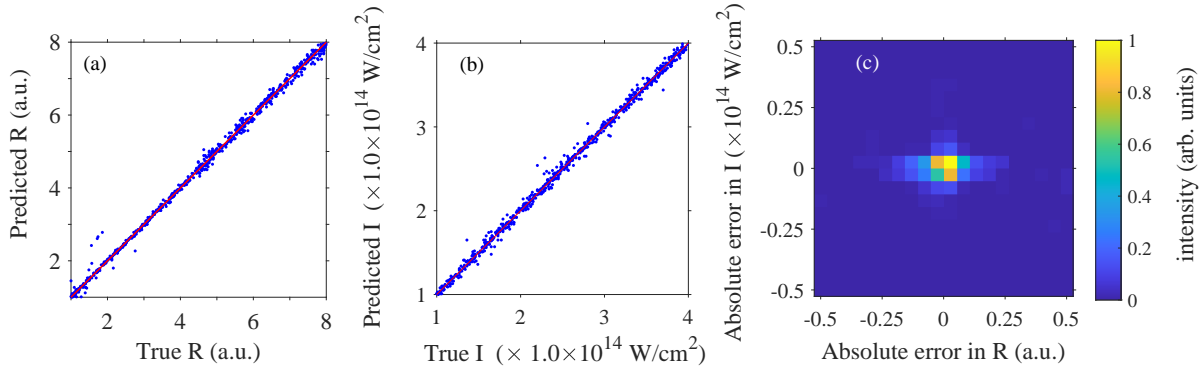} 
\end{center}
\caption{(a) Plot of predicted vs. true internuclear distances in the direct comparison approach. (b) Predicted vs. true laser intensities. (c) The joint 2D histogram of absolute errors in the internuclear distance and laser intensity. The histogram is normalized to unity at the maximum value.}
\label{fig5}
\end{figure}

An approach based on the direct comparison of a given momentum distribution with a set of the precalculated distributions obtained for different internuclear distances was implemented already in Ref.~\cite{Shvetsov2022}. The precalculated set PMDs coincided with the training set used to train the neural network. The MSE was used to compare different images. However, this approach has not been comprehensively compared with the CNN. Only the transferability properties of the direct comparison were discussed in Ref.~\cite{Shvetsov2022}. The results from the direct comparison approach in Fig.~\ref{fig5}(a)-(c) show that the performance is comparable with the neural network. Indeed, for $2250$ images used for comparison and 750 test images, the MAE for the internuclear distance $R$ and the intensity $I$ is equal to $0.06$~a.u. and $0.03\times10^{14}$~W/cm$^2$, respectively. On the other hand, the CNN trained on the same set of $2250$ images to retrieve both the internuclear distance and the laser intensity provides the MAEs for $R$ and $I$ equal to $0.07$~a.u. and $0.05\times10^{14}$~W/cm$^2$, respectively \cite{Shvetsov2022}.

It may appear natural to employ interpolation for producing more PMDs based on the precalculated ones in order to further improve the accuracy of the direct comparison. However, with increasing number of interpolated distributions, a limit in accuracy, with which $R$ and $I$ are retrieved, will be reached. In our case this limit is reached at about $5250$ interpolated PMDs (i.e., $7500$ PMDs used for comparison, including $2250$ distributions of the precalculated set and $5250$ interpolated distributions). The corresponding MAEs for the internuclear distance and laser intensity are $0.055$~a.u. and $0.025\times10^{14}$~W/cm$^2$, respectively. 

The direct comparison approach does not necessarily have to be based on the MSE. Any other suitable measure of similarity between two different images can be applied. Here we show the direct comparison using the SIFT algorithm. Although the SIFT is an involved method, for the sake of completeness, here we briefly discuss its main stages (see Ref.~\cite{Lowe1999} for details). At the first stage of the SIFT the so-called scale-space of a given image is built. The scale-space is produced by the convolution of the given image with the Gaussian kernel $\left(1/2\pi\sigma^2\right)\exp\left[-\left(x^2+y^2\right)/2\sigma^2\right]$ at different scale parameters $\sigma$. The original image is then resized to its half size and the procedure is repeated. As a result, several sets of blurred images (octaves) are produced, and the images of each octave are the same size. Four octaves are usually calculated in the SIFT algorithm. The blurred images are required to produce another set of images called difference of Gaussians. Within each octave the difference of Gaussians is calculated by subtracting the image obtained for the larger value of the parameter $\sigma$ from the neighboring image obtained for the smaller $\sigma$. The resulting sets of images corresponding to different octaves are used to find the key points of the initial image.

At the second stage of the SIFT each pixel of the difference of Gaussians is compared with all $8$ neighboring pixels, as well as with $9$ pixels of the image with larger $\sigma$ and 9 pixels of the image with smaller $\sigma$. The obtained local extrema are the potential key points of the initial image. The generated points that lie along an edge of an initial image or do not have enough contrast are eliminated. The usage of the rest of the points ensures the scale invariance of the SIFT algorithm.

The rotation invariance of the SIFT is provided by the third stage of the algorithm. At this state, a vicinity of each key point is considered and the magnitude and orientation of the gradient in each point of this vicinity is calculated. The obtained values of the magnitude and orientation are used to create a histogram with bins corresponding to different orientation angles. The values that are summed in these bins are the respective gradient magnitudes. The highest peak of the histogram is then chosen, and any other peak exceeding 80\% of the highest one is also used to determine the orientation. As a result, a set of key points with same scale and location but different orientations is found. 

Finally, a feature vector (descriptor) of each key point is calculated using the surrounding pixels. This descriptor is based on the gradient orientation and contains $128$ values. The rotation dependence of the descriptor is excluded by calculating the difference of each gradient orientation and the orientation of the gradient at the key point. The illumination dependence can be eliminated by using a threshold value and renormalization of the feature vector. When two images are compared using SIFT descriptors, the pairwise Euclidean distances between the feature vectors of the first and the second images is to be calculated. Two feature vectors match if the distance between them is less than a chosen threshold value.

\begin{figure}
\begin{center}
\includegraphics[width=.80\textwidth, trim=0 0 10 5]{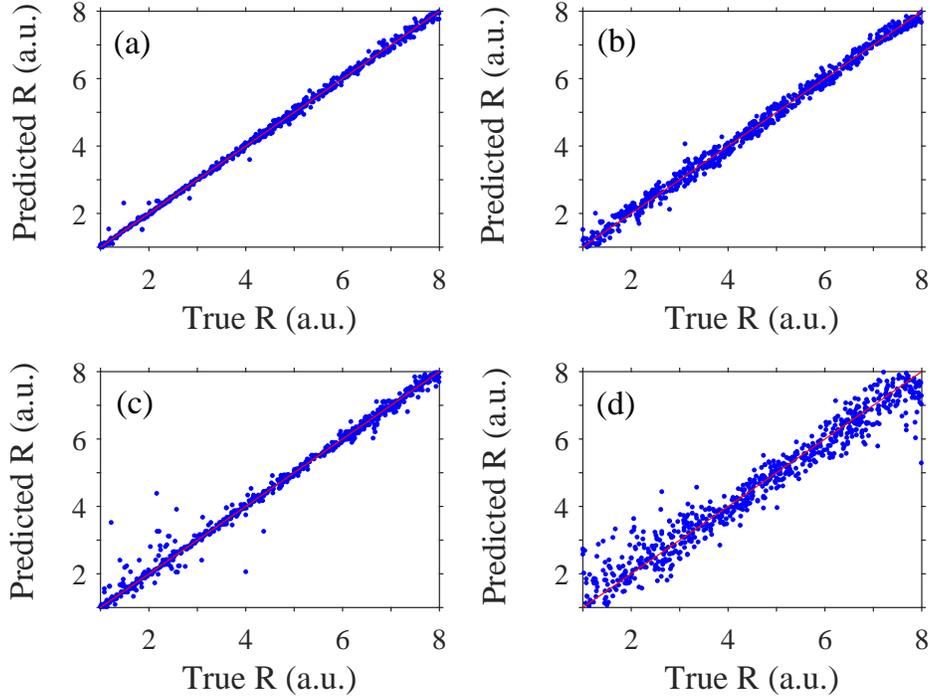} 
\end{center}
\caption{Plots of predicted vs. true internuclear distances illustrating the performance of the alternative approaches to the retrieval of the internuclear distance. Panels (a), (b), (c), and (d) correspond to the direct comparison method employing the SIFT algorithm, the SVM trained using HOGs from different regions of an image, the DT trained on HOGs, and the SVM trained on feature vectors that are extracted with the SIFT algorithm, respectively.}
\label{fig6}
\end{figure}

Therefore, we count numbers of matching feature vectors when comparing a given PMD with the distributions of the precalculated set. The distribution with the maximal number of the matches is used to determine the quantities of interest, i.e., the internuclear distance and the laser intensity. For the sake of brevity, in this example, as in the other following examples, we discuss the reconstruction of the internuclear distance only. The MAE for the internuclear distance $R$ of the direct comparison based on the SIFT algorithm is $0.054$~a.u., see Fig.~\ref{fig6}(a).

However, direct comparison is not the only possible alternative to a neural network. Support vector machines and decision trees can also be used for image classification tasks. For two linearly separable sets of points, the support vector machine (SVM) algorithm finds the hyperplane that corresponds to the maximum distance from it to the nearest data point on each side. Obviously, the hyperplane obtained with the SVM algorithm ensures the largest separation (margin) between the two sets. For an $N$-dimensional space, two sets of points are called linearly separable if there exist at least one $(N-1)$-dimensional hyperplane that separates them. For linearly nonseparable classes, the SVM defines some new space and a mapping that transforms the original space to the new space so that an optimal separating hyperplane exists in this new space. Therefore, in the new space the algorithm seeks for a hyperplane separating the transformed points. The mapping used in the SVM algorithm is implicitly defined by the so-called kernel function. The kernel function replaces the scalar product in the minimization problem solved in the SVM (see, e.g., Ref.~\cite{ScholkopfBook} for details). As a result, the mapping may be nonlinear, and the new space may have more dimensions than the original one.

A decision tree (DT) is a model based on a flowchart-like structure that is used to take decisions. Each internal node of this structure corresponds to some test, and each branch corresponds to the outcome of this test. Finally, each leaf of the tree corresponds to a decision (in our case, a class label). Therefore, the classification rules implemented by a DT are represented by the paths from the root of the tree to its leaves. 

However, to the best of our knowledge, the application of the SVMs and DTs to \textit{image regression} problems is not so well documented. Despite this, we use both these methods to retrieve the internuclear distance from PMDs. As a set of features (attributes) that is needed to train the SVM or DT, we use the histogram of oriented gradients (HOG). It is clear that gradients are efficient for finding the edges and corners of an image. The HOG is calculated in the same way as the histogram in the SIFT algorithm, i.e., the magnitudes of the gradient vectors, whose orientation angles correspond to the same bin, are summed up. The size of the cell we used to extract the HOG is $16\times32$ pixels. The MAEs for the internuclear distance $R$ obtained by using the SVM and DT are equal to $0.13$~a.u. and $0.10$~a.u., respectively [see Figs.~\ref{fig6}(b) and \ref{fig6}(c)]. 

We also combine the SVM with the SIFT algorithm, i.e., we train the SVM using the feature vectors obtained with the SIFT. However, this combination leads to worse results as compared to all other approached discussed here: the corresponding MAE is equal to $0.32$~a.u., see Fig.~\ref{fig6}(b). We attribute this to the fact that for a significant number of PMDs the SIFT algorithm is not able to extract the sufficient number of key points needed to train the SVM. Indeed, for some of the distributions the SIFT algorithm detects only about 15-20 key points. For these reasons, we do not consider the combination of the SIFT with the SVM as an appropriate tool for retrieval of $R$.

Finally, it is interesting to perform transferability tests of the approaches discussed here. To this end, we use the same four data sets as in Sec.~IV. We apply the direct comparison method employing the SIFT descriptors, as well as the SVM and DTs (both trained using the HOG) to these data sets. It should be stressed that in all these cases the precalculated set (or the training set for the SVM and the DT) is the set of the PMDs obtained for fixed laser intensities (i.e., not focal averaged), $1.0 \leq R \leq 8.0$~a.u., $\alpha=0^{\circ}$, and $\varphi=0$. The results of these transferability tests are presented in Table~1. For completeness, in Table~1 we also show the results of the transferability tests for the direct comparison based on the MSE: all the corresponding numbers except the MAE for $R$ in the case of the fluctuating CEP were presented in Ref.~\cite{Shvetsov2022}. We also complement Table~1 with the results obtained by application of the transfer learning technique to the CNN (see Sec.~IV). It is seen that all the alternative methods discussed here possess very little transferability. What is even more important, their potential to become more transferable yet still efficient is rather limited: the precalculated (training) data sets are to be considerably increased. Obviously, this will require a lot of calculations. This is a strong argument for applying neural networks to the problem at hand. 
 
\begin{table}
\caption{\label{table1} The MAE for the internuclear distance $R$ (in a.u.) obtained with the original CNN, the CNN after the application of the transfer learning technique, and by using the alternative methods for four different test sets: focal-volume averaged momentum distributions, distributions for larger internuclear distances, the distributions obtained for nonzero orientation angles, and the PMDs calculated for fluctuating CEP. The data in the first line, except the MAE for nonzero CEPs, were already published in Ref.~\cite{Shvetsov2022}.}  
\begin{ruledtabular}
\begin{tabular}{lcccc}
Method &  Focal averaged & $8.0<R<12.0$~a.u. & $0^{\circ}< \alpha < 180^{\circ}$ & $0<\varphi<2.0\pi$ \\
\hline
CNN & 0.83 & 3.05 & 0.90 & 0.89 \\
CNN + Transfer learning & 0.15 & 0.06 & 0.17 & 0.23 \\
Direct comparison + MSE  & 1.44 & 5.10 & 1.37 & 1.38 \\
Direct comparison + SIFT & 1.19 & 3.57 & 1.14 & 0.71  \\
SVM + HOG & 1.03 & 2.15 & 0.99 & 1.18 \\
DT + HOG & 2.09 & 2.07 & 1.54 & 1.52 \\
\end{tabular}
\end{ruledtabular}
\end{table}

\section{Visualization of the convolutional neural network}

Visualization methods of artificial intelligence are designed to explain and interpret machine learning models. The corresponding new research direction that emerged in the last two decades is often referred to as explainable artificial intelligence. Although the visualization methods are being intensively developed \cite{Atefeh2021,Linardatos2021,Ayyar2021}, there are many open questions in this field of research. Sometimes explanations offered by the visualization methods do not allow to understand what the machine learning model is actually doing. For these reasons, there is even a point of view that it is necessary to stop explaining ``blackbox" machine learning models and use interpretable models instead \cite{Rudin2019}. 

Nevertheless, we apply visualization methods to our CNN. Since the vast majority of the visualization methods deal with the CNN designed for classification of images \cite{Atefeh2021,Ayyar2021}, we train another neural network that classifies the PMDs into the following seven categories: $1.0\leq R <2.0$~a.u., $2.0\leq R< 3.0$~a.u., ...,$7.0\leq R \leq8.0$~a.u. For simplicity, we use only a quarter of each PMD for the training, namely the first quadrant of the $\left(k_x,k_y\right)$ plane. As any classifying neural network (see, e.g., Ref.~\cite{KimBook}), our CNN calculates the probabilities (scores) for a given image to belong to the classes specified above. The classification result is determined by the highest score. The same training and validation data sets (see Sec.~II) are used to train this neural network. We aim to understand what features of an image allow the CNN to assign it to a particular class. 

To achieve this goal, we apply the occlusion sensitivity method, see Ref.~\cite{Zeiler2013}. It is a simple technique that allows us to understand which parts of an image are the most important ones for the CNN to take a classification decision. In this method different parts of the image are occluded with a mask (e.g., gray square) that moves across the image. Simultaneously, the change of the probability score for a specific class is calculated as a function of the occluding mask position. As a result, the occlusion sensitivity map of the image is obtained. This map shows the impact of different parts of the image on the corresponding class score: the parts with high occlusion sensitivity have a positive contribution to the specified class.   

\begin{figure}
\begin{center}
\includegraphics[width=.85\textwidth, trim=20 0 20 0]{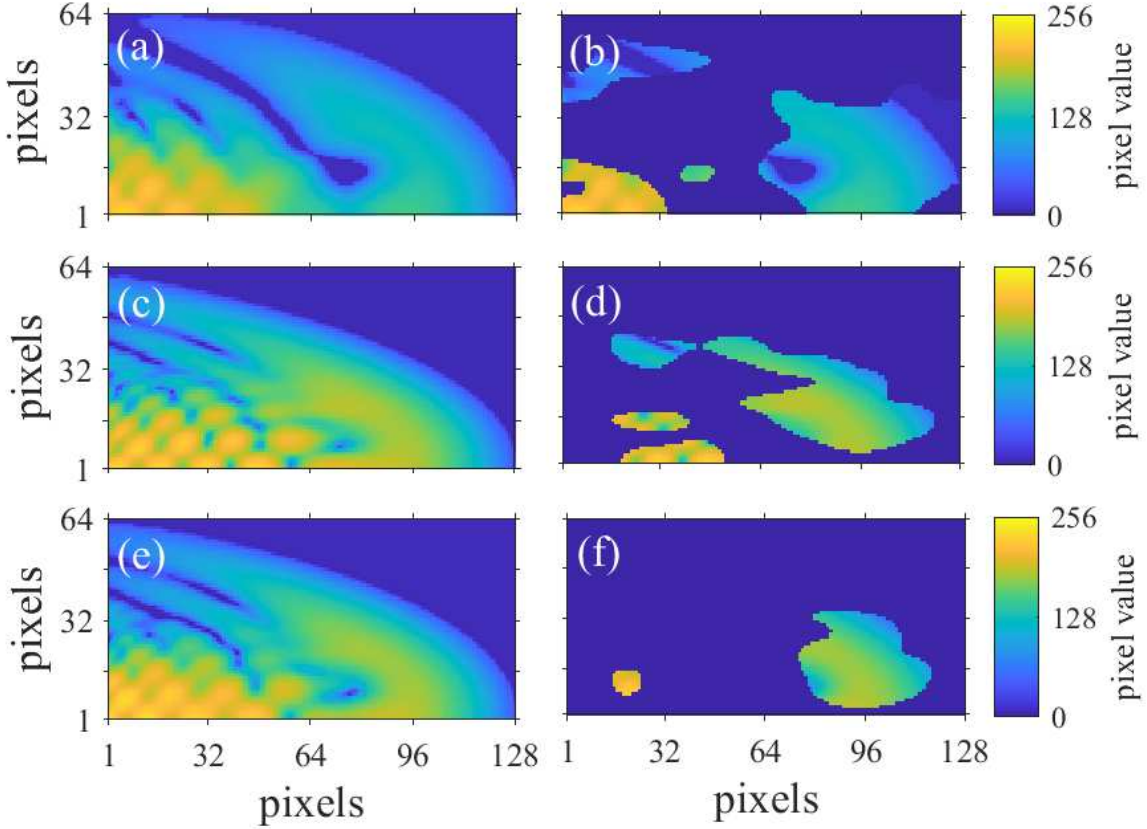} 
\end{center}
\caption{Electron momentum distributions in the first quadrant of the $\left(k_x,k_y\right)$ plane [panels (a),(c), and (e)] and their parts extracted by using the occlusion sensitivity method [panels (b), (d), and (f)] (see text). Panels (a,b), (c,d), and (e,f) correspond to the laser intensities $2.1\times10^{14}$~W/cm$^2$, $2.8\times10^{14}$~W/cm$^2$, and $2.5\times10^{14}$~W/cm$^2$, respectively. The internuclear distances are (a,b) $4.1$~a.u., (c,d) $4.9$~a.u., and (e,f) $4.3$~a.u. The laser pulse duration and the wavelength are as in Fig.~\ref{fig4}, the orientation angle is $\alpha=0^{\circ}$, and the CEP is $\varphi=0$.}
\label{fig7}
\end{figure}

The distributions belonging to the same class, e.g., $4.0\leq R <5.0$~a.u. vary strongly with the peak laser intensity. For this reason, we consider the following intensity intervals: $1.0\times10^{14}\leq I <2.0\times10^{14}$ W/cm$^2$, $2.0\times10^{14}\leq I<3.0\times10^{14}$ W/cm$^2$, and $3.0\times10^{14}\leq I \leq4.0\times10^{14}$ W/cm$^2$ for each of the seven classes. Therefore, we have $7\cdot 3=21$ different alternatives. For each alternative, we randomly choose three PMDs from the validation set and apply the occlusion sensitivity method to these $63$ images. We then extract those parts of the chosen PMDs that correspond to values of the occlusion sensitivity larger than $0.2$. In doing so, we create a ``dictionary" of features corresponding to all seven intervals of the internuclear distance $R$. Examples of chosen PMDs and the corresponding extracted images are shown in Figs.~\ref{fig7}(a)-(f). The number of $63$ images can be further reduced: Many of the resulting images are similar to each other. 

As a test, we use this ``dictionary" to classify the rest of the PMDs of the validation set. Specifically, we directly compare all the images of our ``dictionary" with a given PMD: The image from the ``dictionary" that shows the maximal similarity with the corresponding area (areas) of the momentum distribution of interest determines the class of internuclear distances. Here we use MSE for image comparison. The resulting simple classifier, which is based on the application of the occlusion sensitivity method, shows an accuracy of $67$-$72$\%. Here the accuracy is defined as the ratio of correctly classified momentum dustributions to the whole number of PMDs used for validation. This accuracy varies depending on the set of  distributions that are chosen to extract the relevant features.

The question may arise: What happens if instead of the parts of the PMDs found by the occlusion sensitivity method, we use random parts of the momentum distributions with the same total area as images of the ``dictionary"? From Figs.~\ref{fig7}~(b), (d), and (f) one might speculate that the applicability of our simple classifier is only a consequence of the large sizes of the extracted parts. However, numerical experiments show that randomly chosen parts with the same total area as the joint area of the occlusion-sensitivity features result in a mean classification accuracy of about only $35$\%. The stated number is the mean accuracy over a number of tests, since the accuracy obtained in a single test varies in the range between $20$\% and $46$\% depending on the choice of random areas.

We note that similar classifiers based on alternative visualization methods, namely the gradient-weighted class-activation mapping (Grad-CAM) \cite{Selvaraju2016} and locally interpretable model-agnostic explanations (LIME) \cite{Ribeiro2016}, give correct predictions for 54\% and 59\% of the distributions, respectively. It should be stressed that our primitive classifier is not meant to replace the CNN. Nevertheless, its relatively good performance justifies the use of the occlusion sensitivity method. Since the set of features extracted by occlusion sensitivity can be applied to classify images, these features are indeed characteristic for different ranges of internuclear distance. Therefore, these features can be viewed as the ones used by the neural network when making a classification decision. 

\section{Conclusions and Outlook}

In conclusion, we have investigated a number of problems that arise concerning the application of deep learning for prediction of molecular properties and laser parameters from electron momentum distributions. We have studied the effect of the CEP on the retrieval of the internuclear distance. By using the transfer learning technique, we have made our CNN applicable to the PMDs it was not explicitly trained for: focal volume averaged PMDs and distributions obtained either for internuclear distances, or nonzero angles, or CEPs outside the training range. It is shown that in all these four cases, transfer learning avoids the calculation of large training data sets. These large training sets are needed, if instead of applying the transfer learning technique we train new CNNs. We have shown that transfer learning for focal volume averaged distributions and for distributions obtained for large internuclear distances leads to smaller MAEs as compared with the cases of nonzero orientation angles or CEPs. We attribute this to the fact that focal-volume averaged momentum distributions and the distributions calculated for large internuclear distances are more similar to the PMDs used for training of the initial CNN. Nevertheless, even in the cases of orientation angles and CEPs, the modified CNNs provide MAEs for the internuclear distance of $0.2$~a.u. even though no more than $750$ images used for transfer learning. 

We have compared the usage of the neural network with alternative approaches: direct comparison of the distributions, SVM, and DT. Moreover, when implementing these alternative methods, we have applied not only the MSE (for direct comparison) or HOG (for SVM and DT), but also the more sophisticated SIFT algorithm used in computer vision. Although some of these alternative approaches, e.g., direct comparison in combination with SIFT, can provide better results than those obtained with the CNN, all these methods have very limited transferability. The only possible way to make the alternative methods more transferable is to significantly enlarge the corresponding precalculated data (for direct comparison) or training data (for SVM and DT). Such an increase involves heavy computational costs. 

As a visualization method of machine learning, we have applied the occlusion sensitivity technique to a network trained for the solution of the classification problem: attributing a given momentum distribution to a certain class of internuclear distances. In this way we have extracted the features of the PMDs that allow the CNN to assign a given momentum distribution to one or another class. The appropriateness of the extracted features has been checked by comparing directly a given PMD with the collection of the obtained features, resulting in an accuracy of about 70\%. The application of visualization methods to the CNNs trained for the regression problem rather than the classification problem will be the subject of further studies. Also it will be interesting to apply deep learning for the retrieval of the internuclear distance in case of moving nuclei. Our present results give us reason to believe that deep learning will lead to significant progress in time-resolved molecular imaging including nuclear motion. 
 
\begin{acknowledgements} 
We are grateful to S.~Brennecke, F.~Oppermann, and S.~Yu for stimulating discussions. This work was supported by the Deutsche Forschungsgemeinschaft (Grant No. SH 1145/1-2). 

\end{acknowledgements}

\end{document}